# The response of isotropic composites with viscoelastic matrices

deBotton Gal and Tevet-Deree Limor

The Pearlstone center for aeronautical studies

Department of Mechanical Engineering

Ben-Gurion University

Beer-Sheva 84105, Israel

**Abstract**

Explicit expressions for the behavior of statistically isotropic composites with viscoelastic matrices and linear elastic inclusions are determined by application of the correspondence principle. The behavior of the matrix is linear in dilatation and governed by the standard linear model in shear. We demonstrate that in some cases explicit expressions for the behavior of composites with more general behaviors of the matrix can be derived. We also show that under certain loading conditions the response of the composite may be non-monotonous. This is the result of the interaction between two phases with different viscoelastic behaviors and may appear for more general materials.

## 1. Introduction

In many applications of practical importance the knowledge of the viscoelastic behavior of the composite is essential. Hashin [1] introduced the *correspondence principle* which allows to translate exact expressions for the moduli of linear elastic composites into corresponding expressions for the relaxation moduli of viscoelastic composites with identical microstructures. To determine the variations of the relaxation moduli in the time domain inverse transformations from the Laplace domain must be carried out. Due to their complexity these are usually executed numerically. Nonetheless explicit expressions for the relaxation moduli of fiber composites with viscoelastic matrices whose isochoric behavior is governed by the *standard linear* model were introduced recently by deBotton and Tevet-Deree [2].

In this work we follow the procedure outlined in [2] and consider the class of statistically isotropic composites made out of isotropic viscoelastic matrices with isotropic and linear elastic inclusions. The behavior of the matrix is elastic in dilatation and governed by the standard-linear model in shear. Thus, the shear relaxation modulus of the matrix phase is characterized by a single relaxation time $\tau$ such that

$$\mu(t) = \mu^{(\infty)} + \left(\mu^{(0)} - \mu^{(\infty)}\right)e^{-t/\tau} \qquad (1)$$

Here, $\mu^{(0)}$ is the instantaneous initial modulus at the unstressed and undeformed state. $\mu^{(\infty)}$ is the relaxed (or equilibrium) modulus which characterizes the behavior of the matrix phase under very slow loading conditions. The relative decrease in the shear stiffness of the matrix $g$, is defined such that $\mu^{(0)}g = \mu^{(0)} - \mu^{(\infty)}$. With $g = 1$ (*i.e.*, $\mu^{(\infty)} = 0$) Eq. (1) represents a Maxwell material and with $g \ll 1$ the behavior corresponds to that of a Kelvin-Voigt material. The standard-linear model represents a viscoelastic material that behaves elastically for both sufficiently fast and slow deformations [3].

The expression for the effective bulk relaxation modulus is obtained by utilizing the exact expression of Hashin [4] for the effective bulk modulus of the composite sphere assemblage (CSA). We recall that the expression for the effective bulk modulus of the linear CSA model is identical to the expression for the HS bound [5] on the effective bulk modulus.

The expression for the effective shear relaxation modulus is derived from the HS bound [5] on the shear modulus. Francfort and Murat [6] demonstrated that this bound is optimal by construction of a special sequentially laminated composite which attains the bound. Thus, the result that we obtain can be viewed as an exact expression for the relaxation shear modulus of this sequentially laminated composite. In the context of more general microstructures we recall that [2] considered the class of transversely isotropic composites with viscoelastic isotropic matrices and linear elastic fibers. In [2] the exact results of Hashin and Rosen [7] and Hill [8] and the bounds of Hashin [9] for the class of linear elastic fiber composites were utilized, via the correspondence principle, to determine estimates for the effective viscoelastic behavior of the composite. These estimates were compared with finite element simulations of periodic fiber composites with hexagonal unit cells. Both estimates, the ones based on the exact solutions as well as the ones obtained from the bounds of Hashin [9] provided excellent predictions for the behavior of the viscoelastic periodic composite. Motivated by these findings in this work we determine the expression for the effective relaxation shear modulus which, together with the expression for the bulk modulus, also provide expressions for the effective constrained Young's modulus and the cross relaxation modulus.

## 2. Effective bulk relaxation modulus

The expression for the effective bulk modulus of the linear-elastic CSA composite is [4]

$$\tilde{\kappa}^{(CSA)} = \kappa_1 + c\left(\frac{1}{\kappa_2 - \kappa_1} + \frac{3(1-c)}{3\kappa_1 + 4\mu_1}\right)^{-1}, \tag{2}$$

where $c$ is the inclusions' volume fraction, $\kappa$ is bulk modulus, $\mu$ is shear modulus, and subscripts 1 and 2 correspond to the matrix and the inclusions phases, respectively. We follow the procedure outlined in [2] for applying the correspondence principle and obtain the following expression for the effective bulk relaxation modulus of the composite

$$\tilde{\kappa}^{(CSA)}(t) = \tilde{\kappa}^{(\infty)} + \left(\tilde{\kappa}^{(0)} - \tilde{\kappa}^{(\infty)}\right) e^{-t/\tilde{\tau}_\kappa}, \tag{3}$$

where

$$\tilde{\tau}_\kappa = \frac{4\left(\mu_1^{(0)}/\kappa_1\right) + 3\left[c + (1-c)(\kappa_2/\kappa_1)\right]}{4(1-g)\left(\mu_1^{(0)}/\kappa_1\right) + 3\left[c + (1-c)(\kappa_2/\kappa_1)\right]} \tau. \tag{4}$$

In Eq. (3) the expression for $\tilde{\kappa}^{(0)}$ is given by Eq. (2) with $\mu_1 = \mu_1^{(0)}$. Similarly, the expression for $\tilde{\kappa}^{(\infty)}$ is determined by this equation with $\mu_1 = \mu_1^{(\infty)}$.

From the above expression for $\tilde{\tau}_\kappa$ together with the fact that $0 \leq g \leq 1$ it follows that regardless of the properties of the two constituents and their volume fractions the relaxation time of the composite is strictly larger than the relaxation time of the matrix phase. Also note that $\tilde{\tau}_\kappa$ monotonically increases with $g$. The effective relaxation time $\tilde{\tau}_\kappa$ approaches $\tau$ in the limit of an incompressible matrix (i.e., $\mu_1^{(0)} \ll \kappa_1$) or if the inclusions are markedly stiffer than the matrix (i.e., $\mu_1^{(0)} \ll \kappa_2$). The normalized bulk relaxation modulus can be written as a single term Prony series

$$\tilde{\kappa}^{(CSA)}(t)/\kappa_1 = \left(\tilde{\kappa}^{(0)}/\kappa_1\right)\left[1 - \tilde{g}_\kappa\left(1 - e^{-t/\tilde{\tau}_\kappa}\right)\right], \tag{5}$$

where $\tilde{g}_\kappa$, the relative decrease in the bulk modulus of the composite is

$$\tilde{g}_\kappa \equiv \left(\tilde{\kappa}^{(0)} - \tilde{\kappa}^{(\infty)}\right)/\tilde{\kappa}^{(0)}. \tag{6}$$

Next, we examine two limiting classes of practical importance. The first is of *reinforced composites* where the inclusions are stiffer than the matrix phase. The second consists of *weakened composites* with inclusions that are softer than the matrix. Porous composites belong to the second class.

In the first class we consider the limit $\kappa_2 \gg \kappa_1$. In this case $\tilde{\tau}_\kappa = \tau$ and

$$\tilde{g}_\kappa = \frac{2cg\left(1 - 2\nu_1^{(0)}\right)}{1 + \nu_1^{(0)} + 2c\left(1 - 2\nu_1^{(0)}\right)}, \tag{7}$$

where

$$\nu_1^{(0)} = \frac{3\kappa_1 - 2\mu_1^{(0)}}{6\kappa_1 + 2\mu_1^{(0)}} \tag{8}$$

is the Poisson's ratio of the matrix phase in its unstressed and undeformed state. In the following we restrict our attention to the range $0 \leq \nu_1^{(0)} \leq 0.5$, although materials with negative Poisson's ratio are available [10].

In Fig. 1 the variations of $\tilde{g}_\kappa$ of reinforced composites as functions of $c$ for three values of $\nu_1^{(0)}$ are shown. In all cases $g = 0.95$. Note that $\tilde{g}_\kappa$ monotonically increases with $c$. Clearly, in the dilute limit, where the composite's behavior is dictated by that of the matrix $\tilde{g}_\kappa \to 0$. As $c$ increases the shear stresses in the matrix phase increase resulting in amplification of the viscoelastic character of the composite. On the other hand, $\tilde{g}_\kappa$ decreases with $\nu_1^{(0)}$. Thus, in the limit of an incompressible reinforced matrix $\tilde{g}_\kappa = 0$. As the contrast between the bulk and the shear moduli of the matrix decreases (*i.e.*, smaller $\nu_1^{(0)}$), the viscous deviatoric behavior of the matrix is manifested and hence the perceived decrease in $\tilde{\kappa}^{(\infty)}$. In the limit $c \to 1$ and $\nu_1^{(0)} \to 0$ we have that $\tilde{g}_\kappa \to (2/3)g$. This implies that with an appropriate choice of the composite's constituents and morphology the relative decrease in the effective bulk modulus can become close to that of the matrix phase.

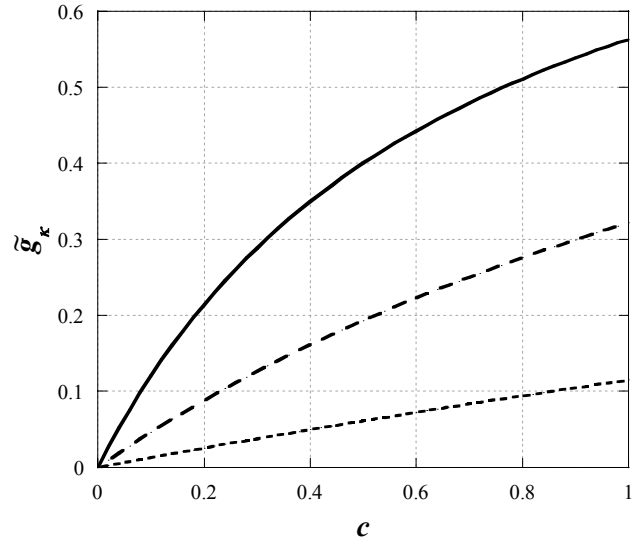

**Fig. 1 The variations of $\tilde{g}_\kappa$ for composites with stiffer inclusions as functions of the inclusions' volume fraction. The continuous, long-dashed and short-dashed curves are for $\nu_1^{(0)} = 0.1$, $0.33$ and $0.45$, respectively. For all curves $g = 0.95$.**

In the second case, of viscoelastic matrices weakened by softer inclusions (*e.g.*, porous composites) we consider the limit $\kappa_1 \gg \kappa_2$. In this limit

$$\widetilde{\tau}_\kappa = \frac{2(1-2v_1^{(0)}) + c(1+v_1^{(0)})}{2(1-g)(1-2v_1^{(0)}) + c(1+v_1^{(0)})} \tau. \tag{9}$$

The effective relaxation time of these composites can become very large as the $c$ decreases. The fastest response time is attained in the limit of an incompressible matrix ($v_1^{(0)} \to 0.5$).

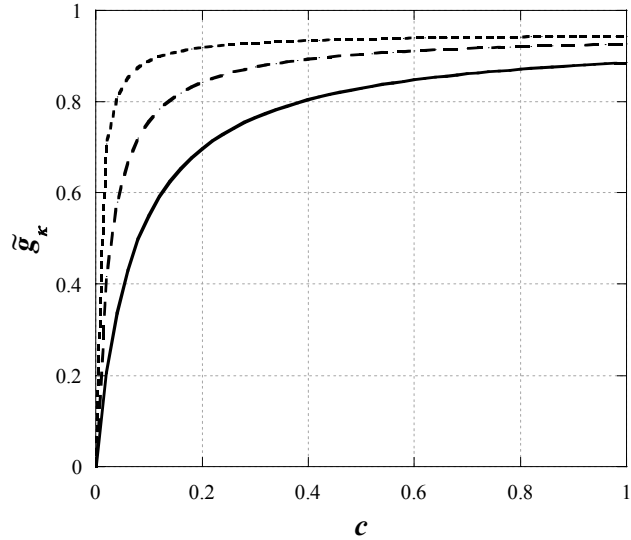

**Fig. 2 The variations of $\widetilde{g}_\kappa$ for composites with softer inclusions as functions of the inclusions' volume fraction. The continuous, long-dashed and short-dashed curves are for $v_1^{(0)} = 0.1$, $0.33$ and $0.45$, respectively. For all curves $g = 0.95$.**

In this limit of weakened composites the expression for $\widetilde{g}_\kappa$ reduces to

$$\widetilde{g}_\kappa = \frac{cg(1+v_1^{(0)})}{2(1-g)(1-2v_1^{(0)}) + c(1+v_1^{(0)})}. \tag{10}$$

The variations of $\widetilde{g}_\kappa$ as functions of $c$ for three values of $v_1^{(0)}$ are shown in Fig. 2. In contrast with the tendency of $\widetilde{\tau}_\kappa$, $\widetilde{g}_\kappa$ increases with $c$. Thus, as the volume fraction of the soft inclusions increases, simultaneously the relaxation time decreases and the softening of the bulk modulus is increasing. In particular, we note that as $\kappa_2/\kappa_1 \to 0$ and $\kappa_1 \gg \mu_1$ the expression for the effective bulk modulus of a linear elastic CSA is

$$\widetilde{\kappa}^{(CSA)} \to \frac{4(1-c)}{3c} \mu_1. \tag{11}$$

Due to the linearity of the Laplace transform it follows that the expression for the effective bulk relaxation modulus of a CSA with soft inclusions and a nearly incompressible matrix with an *arbitrary* relaxation function $\mu_1(t)$ is

$$\tilde{\kappa}^{(CSA)}(t) = \frac{4(1-c)}{3c}\mu_1(t). \tag{12}$$

Eq. (12) is the inverse of the corresponding relation determine by Hashin [1] for the creep compliance of a porous composite in the limit of incompressible material. We note that any available experimental data for the matrix phase can be immediately transferred into a data concerning the bulk behavior of the weakened composite. Eq. (12) implies that the bulk behavior of the composite is dictated by the shear relaxation modulus of the matrix.

## 3. Effective shear relaxation modulus

The HS bound [5] for the shear modulus of a linear-elastic isotropic composite is

$$\tilde{\mu}^{(HS)} = \mu_1 + c\left(\frac{1}{\mu_2 - \mu_1} + \frac{2(1-c)(\kappa_1 + 2\mu_1)}{5\mu_1(\kappa_1 + (4/3)\mu_1)}\right)^{-1}. \tag{13}$$

Here, $\tilde{\mu}^{(HS)}$ is a lower bound if phase 2 is stiffer than phase 1 (*e.g.* $\mu_2 \geq \mu_1$ and $\kappa_2 \geq \kappa_1$) and an upper bound in the opposite case. Eq. (13) may be rewritten in the form

$$\tilde{\mu}^{(HS)} = R\mu_1 \frac{1 + Q_1 + Q_0}{1 + q_1 + q_0}, \tag{14}$$

where

$$R = \frac{2(1-c)}{2+3c},$$

$$Q_1 = \frac{(3+2c)\mu_2 + (9/4)(1-c)\kappa_1}{2(1-c)}\left(\frac{1}{\mu_1}\right),$$

$$Q_0 = \frac{(3/4)(2+3c)\mu_2\kappa_1}{2(1-c)}\left(\frac{1}{\mu_1}\right)^2,$$

$$q_1 = \frac{3(1-c)\mu_2 + (3/4)(3+2c)\kappa_1}{2+3c}\left(\frac{1}{\mu_1}\right),$$

and

$$q_0 = \frac{(3/2)(1-c)\mu_2\kappa_1}{2+3c}\left(\frac{1}{\mu_1}\right)^2.$$

In passing, we define for later reference

$$\hat{m}_1 = (1/2)\left(q_1 + \sqrt{q_1^2 - 4q_0}\right),$$
$$\hat{m}_2 = (1/2)\left(q_1 - \sqrt{q_1^2 - 4q_0}\right). \tag{15}$$

The term inside the square roots in Eq. (15) is

$$\frac{\left(3(1-c)\mu_2 - (3/4)(3+2c)\kappa_1\right)^2 + 15(1-c)\mu_2\kappa_1}{(2+3c)^2}\left(\frac{1}{\mu_1}\right)^2,$$

and hence $\hat{m}_1$ and $\hat{m}_2$ are real and positive.

Applying the correspondence principle, with the behavior of the matrix phase governed by Eq. (1), we obtain an expression for the effective shear relaxation modulus

$$\tilde{\mu}^{(HS)}(t) = \tilde{\mu}^{(\infty)} + \mu_1^{(0)} gR\, e^{-t/\tau} + (1/2)\left(\tilde{\mu}^{(0)} - \tilde{\mu}^{(\infty)} - \mu_1^{(0)} gR(1-\Delta)\right) e^{-t/\tilde{\tau}_1}$$
$$+ (1/2)\left(\tilde{\mu}^{(0)} - \tilde{\mu}^{(\infty)} - \mu_1^{(0)} gR(1+\Delta)\right) e^{-t/\tilde{\tau}_2}. \tag{16}$$

Here the composites' effective relaxation times are

$$\tilde{\tau}_1 = \frac{1+\hat{m}_1}{1-g+\hat{m}_1}\tau,$$
$$\tilde{\tau}_2 = \frac{1+\hat{m}_2}{1-g+\hat{m}_2}\tau,$$

and

$$\Delta = \frac{(2-g)\Delta_1 + (1-g)\Delta_2}{(\hat{m}_2 - \hat{m}_1)(1+\hat{m}_1)(1+\hat{m}_2)(1-g+\hat{m}_1)(1-g+\hat{m}_2)},$$

where

$$\Delta_1 = (q_1 - 2(1+Q_1))\, q_0^2 + (2Q_0 + (q_1 + Q_0 - Q_1)\, q_1)\, q_0,$$
$$\Delta_2 = (2Q_1 - q_1)\, q_0^2 + (2Q_1 - (3+Q_0)\, q_1)\, q_0 + (Q_0 + (q_1 - Q_1)\, q_1)\, q_1,$$

and $\tilde{\mu}^{(0)}$ and $\tilde{\mu}^{(\infty)}$ are evaluated from Eq. (13) with $\mu_1 = \mu_1^{(0)}$ and $\mu_1 = \mu_1^{(\infty)}$, respectively. $\tilde{\tau}_1$, $\tilde{\tau}_2$ and $\Delta$ are evaluated with $\mu_1 = \mu_1^{(0)}$ in Eqs. (14-15). We note that since $\hat{m}_1 > \hat{m}_2$ then $\tilde{\tau}_2 > \tilde{\tau}_1$ and both are strictly larger than $\tau$. Also note that both increase with $g$. However, while $\tilde{\tau}_1$ approaches $\tau$ as the contrast between the phases increases, $\tilde{\tau}_2$ is strictly larger than $\tau$.

Eq. (16) can be normalized by the initial shear modulus of the matrix phase and rewritten as a three terms Prony series

$$\tilde{\mu}^{(HS)}(t)/\mu_1^{(0)} = \left(\tilde{\mu}^{(0)}/\mu_1^{(0)}\right)\left[1 - \left(\tilde{g}(1-e^{-t/\tau}) + \tilde{g}_1(1-e^{-t/\tilde{\tau}_1}) + \tilde{g}_2(1-e^{-t/\tilde{\tau}_2})\right)\right], \tag{17}$$

where

$$\tilde{g} = gR\,\mu_1^{(0)}/\tilde{\mu}^{(0)},$$

$$\tilde{g}_1 = (1/2)[\tilde{g}_\mu - \tilde{g}(1-\Delta)],$$

$$\tilde{g}_2 = (1/2)[\tilde{g}_\mu - \tilde{g}(1+\Delta)],$$

and the relative decrease in the effective shear modulus of the composite is

$$\tilde{g}_\mu \equiv \left(\tilde{\mu}^{(0)} - \tilde{\mu}^{(\infty)}\right)/\tilde{\mu}^{(0)} = \tilde{g} + \tilde{g}_1 + \tilde{g}_2. \tag{18}$$

Once again we consider the classes of reinforced and weakened composites. In the first case we assume that $\mu_2 \gg \mu_1^{(0)}$. Hence, the effective relaxation time $\tilde{\tau}_1$ approaches that of the matrix phase (*i.e.* $\tilde{\tau}_1 = \tau$) and

$$\tilde{\tau}_2 = \frac{4 - 5\nu_1^{(0)}}{4 - 5\nu_1^{(0)} - 3g(1 - 2\nu_1^{(0)})}\,\tau. \tag{19}$$

Interestingly, we note that $\tilde{\tau}_2$ is independent of $c$ and decreases with $\nu_1^{(0)}$. Only in the limit of an incompressible matrix it approaches $\tau$. The effective shear relaxation modulus reduces to

$$\tilde{\mu}^{(HS)}(t)/\mu_1^{(0)} = \left(\tilde{\mu}^{(0)}/\mu_1^{(0)}\right)\left[1 - \left(\tilde{g}(1 - e^{-t/\tau}) + \tilde{g}_2(1 - e^{-t/\tilde{\tau}_2})\right)\right], \tag{20}$$

where

$$\tilde{g} = \frac{2g(3 + 2c)(4 - 5\nu_1^{(0)})}{3(8 + 7c - 5\nu_1^{(0)}(2 + c))},$$

and

$$\tilde{g}_2 = \frac{5cg(1 + \nu_1^{(0)})^2}{3(8 + 7c - 5\nu_1^{(0)}(2 + c))(4 - 5\nu_1^{(0)} - 3g(1 - 2\nu_1^{(0)}))}.$$

The variations of $\tilde{g}$ and $\tilde{g}_2$ of reinforced composites with $g = 0.95$ as functions of $c$ for three values of $\nu_1^{(0)}$ are shown in Fig. 3. The trends of $\tilde{g}$ and $\tilde{g}_2$ are opposite. $\tilde{g}$ decreases with $\nu_1^{(0)}$ and $c$, while $\tilde{g}_2$ increases with both. If further the matrix is incompressible, as was noted by Hashin [1] the effective shear relaxation modulus approaches that of the matrix phase (*i.e.,* $\tilde{\mu}^{(HS)}(t) \approx \mu_1(t)$).

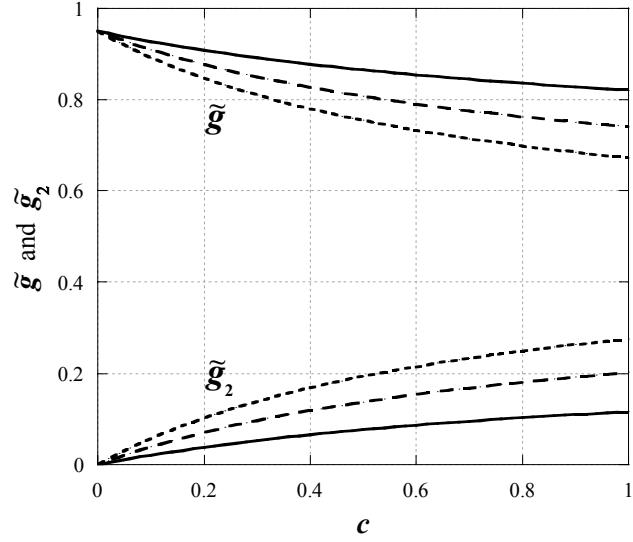

**Fig. 3** The variations of $\tilde{g}$ and $\tilde{g}_2$ for composites with stiffer inclusions as functions of the inclusions' volume fraction. The continuous, long-dashed and short-dashed curves are for $v_1^{(0)} = 0.1$, $0.33$ and $0.45$, respectively. For all curves $g = 0.95$.

In the limit of soft inclusions in viscoelastic matrices or porous composites, we assume that $\mu_1^{(0)} \gg \mu_2$. For this case $\tilde{g}_2 = 0$, and the expression for the effective shear relaxation modulus reduces to

$$\tilde{\mu}^{(HS)}(t)/\mu_1^{(0)} = \left(\tilde{\mu}^{(0)}/\mu_1^{(0)}\right)\left[1 - \left(\tilde{g}\left(1 - e^{-t/\tau}\right) + \tilde{g}_1\left(1 - e^{-t/\tilde{\tau}_1}\right)\right)\right], \qquad (21)$$

where

$$\tilde{\tau}_1 = \frac{2(2+3c)(1-2v_1^{(0)}) + (3+2c)(1+v_1^{(0)})}{2(1-g)(2+3c)(1-2v_1^{(0)}) + (3+2c)(1+v_1^{(0)})}\tau,$$

$$\tilde{g} = \frac{2g\left[2c(5v_1^{(0)} - 4) + (5v_1^{(0)} - 7)\right]}{(2+3c)(5v_1^{(0)} - 7)},$$

and

$$\tilde{g}_1 = \frac{5cg(3+2c)(1+v_1^{(0)})^2}{(2+3c)(5v_1^{(0)} - 7)\left[2g(2+3c)(1-2v_1^{(0)}) + 2c(5v_1^{(0)} - 4) + (5v_1^{(0)} - 7)\right]}.$$

The effective relaxation time $\tilde{\tau}_1$ decreases with $v_1^{(0)}$ and increases with $c$. In the limit of incompressible matrix it approaches $\tau$. In Fig. 4 the variations of $\tilde{g}_1$ and $\tilde{g}$ as functions of $c$ for three values of $v_1^{(0)}$ are shown. The trends of $\tilde{g}$ and $\tilde{g}_1$ are opposite.

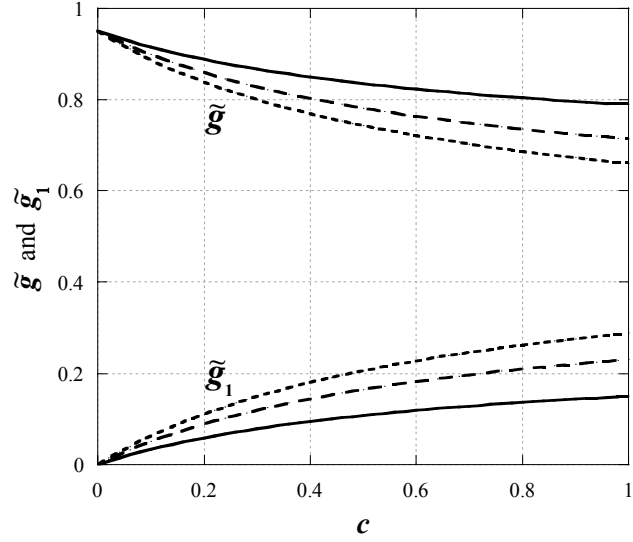

**Fig. 4 The variations of $\widetilde{g}$ and $\widetilde{g}_l$ for composites with softer inclusions as functions of the inclusions' volume fraction. The continuous, long-dashed and short-dashed curves are for $v_1^{(0)} = 0.1$, $0.33$ and $0.45$, respectively. For all curves $g = 0.95$.**

## 4. Effective constrained Young's relaxation modulus

The constrained Young's modulus $n$ relates the longitudinal stress and strain in a state of one-dimensional straining. For linear-elastic and isotropic materials $n = \kappa + (4/3)\mu$. Hence, from the linearity of the Laplace transform for the statistically isotropic viscoelastic composite

$$\widetilde{n}(t) = \widetilde{\kappa}(t) + (4/3)\widetilde{\mu}(t), \tag{22}$$

where the expressions for $\widetilde{\kappa}(t)$ and $\widetilde{\mu}(t)$ are given in Eqs. (3) and (16), respectively. Accordingly, the expression for $\widetilde{n}(t)$ may be written in terms of a four terms Prony series.

## 5. Effective cross relaxation modulus

The cross-modulus $l$ relates the uniaxial stress response to the lateral straining and vise versa. For a linear-elastic isotropic material $l = \kappa - (2/3)\mu$ and due to the linearity of the Laplace transform the effective relaxation cross-modulus of the composite is

$$\widetilde{l}(t) = \widetilde{\kappa}(t) - (2/3)\widetilde{\mu}(t), \tag{23}$$

where $\widetilde{\kappa}(t)$ and $\widetilde{\mu}(t)$ are given in Eqs. (3) and (16). Accordingly, the relative variation in the effective cross-modulus of the composite is

$$\widetilde{g}_l \equiv \left(\widetilde{l}^{(0)} - \widetilde{l}^{(\infty)}\right)/\widetilde{l}^{(0)} = \left(\widetilde{\kappa}^{(0)}\widetilde{g}_\kappa - (2/3)\widetilde{\mu}^{(0)}\widetilde{g}_\mu\right)/\widetilde{l}^{(0)}, \tag{24}$$

where $\widetilde{g}_\kappa$ and $\widetilde{g}_\mu$ are given in Eqs. (6) and (18), respectively.

In contrast with the relaxation cross-modulus of the matrix $l_1(t) = \kappa_1 - (2/3)\mu_1(t)$ which is a monotonically increasing function of time, or the other effective relaxation moduli of the composite which are monotonically decreasing functions of time, $\widetilde{l}(t)$ might monotonically increase, decrease or be a non-monotonous function of time. This results from the different signs of the exponential terms and depends on the properties of the constituents and their concentrations. Moreover, we note that $\widetilde{l}^{(\infty)}$ can be larger than, smaller than or equal to $\widetilde{l}^{(0)}$. When $\widetilde{l}^{(\infty)} \approx \widetilde{l}^{(0)}$ it is possible that initially $\widetilde{l}(t)$ increases and then decreases to $\widetilde{l}^{(\infty)}$ or vise versa. According to Eq. (24) $\widetilde{l}(t)$ behaves non-monotonically when

$$\widetilde{\kappa}^{(0)}\widetilde{g}_\kappa \approx (2/3)\widetilde{\mu}^{(0)}\widetilde{g}_\mu. \qquad (25)$$

To highlight this interesting phenomenon we examine two composites whose effective relaxation moduli satisfy Eq. (25). As in the previous subsections we consider composites with stiff and with soft inclusions. For both composites we assume for the viscoelastic matrix phase $g = 0.95$, $v_1^{(0)} = 0.33$ and inclusions' volume fraction $c = 0.4$. The contrasts in the phases moduli of the reinforced composite are $\kappa_2/\kappa_1 = 20$ and $\mu_2/\mu_1^{(0)} = 2$, and for the weakened composite $\kappa_2/\kappa_1 = 0.3$ and $\mu_2/\mu_1^{(0)} = 0.2$. The non-monotonous temporal variations of $\widetilde{l}(t)/\widetilde{l}^{(0)}$ for the two composites are shown in Fig. 5.

As was noted in [2] this non-monotonous behavior occurs due to the interaction between two phases with different viscoelastic responses. Thus, in a homogeneous body whose isochoric behavior is viscoelastic and is subjected to a uniform lateral compressive displacement boundary condition, the transverse stresses relax with time while the longitudinal stresses increase. This is also true for the viscoelastic matrix phase in the composite. However, in the composite, as the transverse stresses in the matrix relax, the linear-elastic inclusions expand in the transverse plane. Consequently, the normal stresses in the inclusions relax too, and particularly, the longitudinal stress decreases. Thus, in the composite there are two competing mechanisms that influence the total stress in the longitudinal direction. The increasing stress in the matrix and the decreasing stress in the inclusion. Depending on the properties of the two constituents and their concentrations, the overall longitudinal stress in the composite will either increase, decrease or vary in a non-monotonous manner.

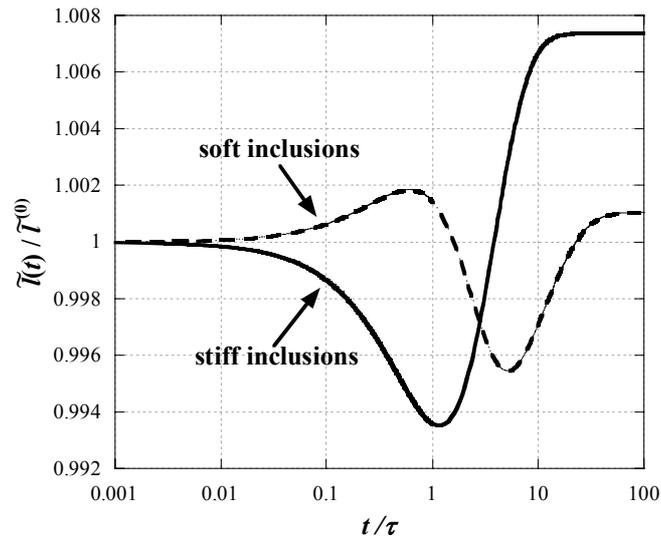

**Fig. 5:** The temporal variations of the effective relaxation cross-modulus normalized by $\tilde{l}^{(0)}$.

**Acknowledgment**

This work was supported by the Israel Science Foundation founded by the Israel Academy of Sciences and Humanities (Grant No. 317/99-2).

**References**

[1] Hashin, Z., 1965, "Viscoelastic behavior of heterogeneous media," J. Appl. Mech. Trans. ASME, **32**, pp. 630–636.
[2] deBotton, G., and Tevet-Deree, L., 2004, "The response of a fiber-reinforced composite with a viscous matrix phase," J. Composite Materials, **38**, pp. 1255–1277.
[3] Banks-Sills, L., and Benveniste, Y., 1983, "Steady interface crack propagation between two viscoelastic standard solids," Int. J. Fracture, **21**, pp. 243–260.
[4] Hashin, Z., 1962, "The elastic moduli of heterogeneous materials," J. Appl. Mech. Trans. ASME, **29**, pp. 143–150.
[5] Hashin, Z., and Shtrikman, S., 1963, "A variational approach to the theory of the elastic behaviour of multiphase materials," J. Mech. Phys. Solids, **11**, pp. 127–140.
[6] Francfort, G., and Murat, F., 1986, "Homogenization and optimal bounds in linear elasticity," Arch. Rational. Mech. Anal., **94**, pp. 307–334.
[7] Hashin, Z., and Rosen, B.W., 1964, "The elastic moduli of fiber-reinforced materials," J. Appl. Mech., Trans. ASME, **31**, pp. 223–232.
[8] Hill, R., 1964, "Theory of mechanical properties of fiber-strengthened materials: I. elastic behaviour," J. Mech. Phys. Solids, **12**, pp. 199–212.
[9] Hashin, Z., 1965, "On elastic behaviour of fibre reinforced materials of arbitrary transverse phase geometry," J. Mech. Phys. Solids, **13**, pp. 119–134.
[10] Lakes, R.S., 1991, "The time-dependent poisson's ratio of viscoelastic cellular materials can increase or decrease," Cellular Polymers, **10**, pp. 466–469.